\documentclass[10pt, onecolumn]{IEEEtran}
\usepackage[utf8]{inputenc}
\usepackage{amsmath}
\usepackage{amsfonts}
\usepackage{amssymb}
\usepackage{graphics}
\usepackage{graphicx}

\newcommand{\Rmnum}[1]{\MakeUppercase{\romannumeral #1}}

\begin{document}

\title{An Analysis on the Inter-Cell Station Dependency Probability in an IEEE~802.11 Infrastructure WLANs}

\author{
	\IEEEauthorblockN{ Albert Sunny\IEEEauthorrefmark{3}, Joy Kuri\IEEEauthorrefmark{3}, Anurag Kumar}\\
	\IEEEauthorblockA{Department of Electrical Communication Engineering, Indian Institute of Science, Bangalore, India\\
    \IEEEauthorrefmark{3}Department of Electronic Systems Engineering, Indian Institute of Science, Bangalore, India} \\
	Email: salbert@dese.iisc.ernet.in, kuri@cedt.iisc.ernet.in, anurag@ece.iisc.ernet.in}

\maketitle

\section{System Model} \label{sec:system_model}
To ensure desired rate (say atleast $r_t$ Mbps) of association to wireless devices in IEEE 802.11g infrastructure mode, we may have to deploy a dense layout of access points, with significant overlaps among their coverage regions. Let $P_t$ be the power level required to ensure a target rate of at least $r_t \, Mbps$. 
\begin{figure}[ht]
\centering
\includegraphics[scale=0.5]{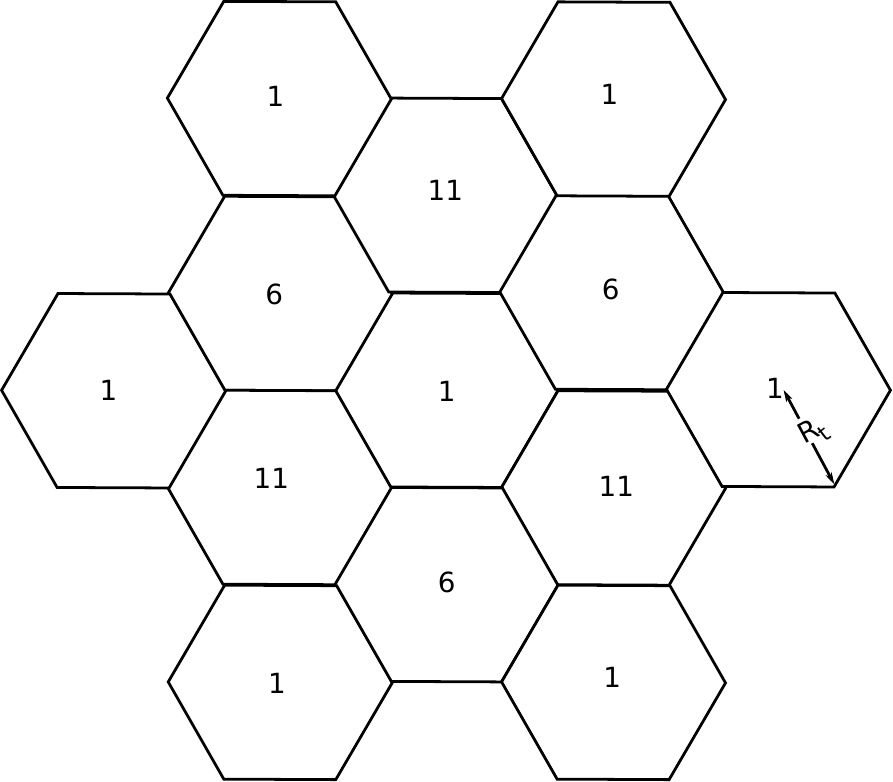}
\caption{Hexagonal Micro-cellular layout of IEEE 802.11g with cell radius $R_t$ and $3$ non-overlapping channels.}
\label{fig:hlayout}
\end{figure}

\noindent
The cell radius $R_t$ is related to the power level $P_t$ as follows
\begin{equation} \label{eq:path_loss_r_t}
P_t = S \cdot 10^{\frac{-\xi}{10}} \cdot \left( \frac{R_t}{R_0} \right)^{- \eta}
\end{equation}
where $S$ is the transmit power, $R_0$ is the ``far field'' \emph{reference distance}, $\eta$ is the path loss exponent and $\xi$ is a Gaussian random variable with mean $0$ and variance $\sigma^2$. Let $r_{min}$ be the minimum transmission rate possible. Let $P_{min}$ and $R_{min}$ be the power level and distance at which $r_{min}$ is sustainable. Then, 
\begin{equation} \label{eq:path_loss_r_min}
P_{min} = S \cdot 10^{\frac{-\xi}{10}} \cdot \left( \frac{R_{min}}{R_0} \right)^{- \eta}
\end{equation}
Dividing Equation \eqref{eq:path_loss_r_min} by Equation \eqref{eq:path_loss_r_t} and rearranging the terms, we obtain
$$R_{min} = R_t \cdot \left( \frac{P_t}{P_{min}}\right)^{1/\eta}$$

Let $R_{cs}$ and $R_i$ be the carrier sensing and interference range of a node in a cell respectively. Let us assume that We assume $R_i = \alpha \cdot R_{cs}$, where $\alpha \geq 1$  which implies that any node within the $R_{cs}$ of a receiver can cause interference and nodes outside the $R_{cs}$ cannot. Thus, the interference region of a node is a disk of radius $R_{cs} = \gamma(\eta, P_t, P_{min}) \cdot R_t$ centred at the node itself, where $\gamma(\eta, P_t, P_{min}) = 2 \alpha \cdot \left( \frac{P_t}{P_{min}}\right)^{1/\eta}$. For ease of analysis, we assume that the interference region is a regular hexagon inscribed with the disk of radius $R_{cs}$. 

Given an association of stations with access points, we can think of each STA-AP association as a \emph{link}. For two stations associated with two different access points, we say that the corresponding links are \emph{dependent} if throughputs obtained on each link with simultaneous bulk TCP transfers are lower than the throughputs obtained when the transfers are performed on the links individually. Based on the above definition of \emph{dependence}, it is easy to see that two stations associated with the same access point are \emph{dependent}. This dependence relation between links is represented as an undirected graph called the \emph{link dependence} graph, whose vertex set is the set of links.

In this document, we are primarily interested in computing the probabilities of various types of dependencies that can occur in a deployment as shown in Figure \ref{fig:hlayout}. Let us assume that we have are given a dependence graph $\mathcal{G}$. Recall that nodes in the dependence graph corresponds to link the actual network (a STA-AP association). Also, since each station can associate with only one access point at a time, each station will correspond to one and only one link in the dependence graph. Thus, we can correlate a link in the dependence graph to dependency between two stations. Consider two stations $S_1$ and $S_2$. Let the stations $S_1$ and $S_2$ be associated with access point $A_1$ and $A_2$ respectively. We classify the dependency between stations $S_1$ and $S_2$ into three main types as follows:  
 
\begin{itemize}
\item \textit{\textbf{{Type \Rmnum{1}} Dependence}}: Stations $S_1$ and $S_2$ are within the interference range of each other. The stations are also outside the interference range of the each others access point and the access points do not interfere with each other.
\item \textit{\textbf{{Type \Rmnum{2}} Dependence}}: Access points of stations $S_1$ and $S_2$ interfere with each other.
\item \textit{\textbf{{Type \Rmnum{3}} Dependence}}: Access points of stations $S_1$ and $S_2$ do not interfere with each other. Station $S_2$ is within the interference range of access points $A_1$ and/or station $S_1$ is within the interference range of access points $A_2$.
\end{itemize}

Let $D_j = \nu_j \cdot R_t$ be the distance between the centres of the $j^{\textrm{th}}$ tier co-channel cells. The maximum and minimum distance between two nodes in the $j^{\textrm{th}}$ tier co-channel cells are $D_j + 2R_t$ and $D_j - 2R_t$ respectively. Thus, if stations $S_1$ and $S_2$ belong to cells whose centres are atleast $R_i + 2R_t$ units apart, the stations will not depend on each other. Let 
$$j_0 = \max \left \{ j \geq 0 \, : \, \nu_j < \gamma(\eta, P_t, P_{min}) + 2 \right \}$$

Now, consider a station (say station $S_1$). Let $A_1$ be the access point to which this station is associated with. Let station $S_1$ belong to a hexagonal cell $d \in \mathcal{N}$, where $\mathcal{N}$ denotes the collection of cells deployed as in Figure \ref{fig:hlayout}. Now, we consider the hexagonal cell of radius $\nu_{j_0} \cdot R_t$ centred at cell $d$ and try to compute the dependency probabilities of another station ($S_2$), when $S_2$ is present within the hexagon of radius $\nu_{j_0} \cdot R_t$ on the same channel as station $S_1$.

\section{Probability of various dependencies}

\subsection{Probability of Type \Rmnum{1} dependency} \label{sec:type_1}
In this section, we find the probability of type \Rmnum{1} dependency between two station $S_1$ and $S_2$. Let $p^{(1)}_j$ denote the probability that two stations have type \Rmnum{1} dependency, given that they belong to co-channels cells with centres  $D_j$ unit apart.

When the cells are partially dependent, to ensure type \Rmnum{1} dependency between $S_1$ and $S_2$, we require $S_1$ and $S_2$  to be in the interference range of each other. We also require $S_1$ and $S_2$  be outside the interference range access points $A_2$ and $A_1$ respectively. 

Let  $\mathcal{H}((x,y),R)$ denote a regular hexagon of radius $R$ centred at $(x,y)$. Let the positions of the access points $A_1$ and $A_2$ be $(x_1,y_1)$ and $(x_2,y_2)$ respectively. Now, let us define the following
\begin{eqnarray*}
\Delta^j_1 = \{ (x,y) \in \mathbb{R}^2 : (x,y) \in  \mathcal{H}((x_1,y_1),R_t),  (x,y) \notin  \mathcal{H}((x_2,y_2),R_{i}), \, 
A_1 \textrm{ and } A_2 \textrm{ are in } j^{\textrm{th}} \textrm{ tier co-channel cells} \} \\
\Delta^j_2 = \{ (x,y) \in \mathbb{R}^2 : (x,y) \in  \mathcal{H}((x_2,y_2),R_t),  (x,y) \notin  \mathcal{H}((x_1,y_1),R_{i}), \, 
A_1 \textrm{ and } A_2 \textrm{ are in } j^{\textrm{th}} \textrm{ tier co-channel cells} \}
\end{eqnarray*}
i.e.,  $\Delta^j_1/\Delta^j_2$ denotes the area outside the interference range of access point $A_2/A_1$ and within the hexagonal cell of radius $R_t$ centred at access point $A_1/A_2$, when $A_1$ and $A_2$ are $j^{\textrm{th}}$ tier co-channel cells. Let $p_1(x,y)$ denote the probability that $S_1$ is at $(x,y) \in \Delta^j_1$ . Also, let $p_2(x,y)$ denote the probability that $S_2 \in \mathcal{H}((x,y),R_{i}) \cap \Delta^j_2$, given that $S_1$ is at $(x,y)$. Now, we can compute the probability of type \Rmnum{1} dependency between stations $S_1$ and $S_2$ as
\begin{equation} \label{eq:integral}
p^{(1)}_j = \int_{\Delta^j_1} p_1(x,y) \cdot p_2(x,y) \, dx \, dy
\end{equation}

The stations are uniformly distributed within the regular hexagonal cell with radius $R_t$ of their associated access points. Thus, we have $p_1(x,y) = \frac{1}{3\sqrt{3}R_t^2/2}$. Since station $S_2$ also has a uniform distribution within its associated hexagonal cell, we have
$$p_2(x,y) =  \frac{Area(\mathcal{H}((x,y),R_{i}) \cap \Delta^j_2)}{3\sqrt{3}R_t^2/2}$$
where $Area(\mathcal{X}), \mathcal{X} \subset \mathbb{R}^2$ denotes the area of the region in $\mathbb{R}^2$ denoted by the set $\mathcal{X}$. Thus, integral \eqref{eq:integral} reduces to
\begin{equation} \label{eq:integral_final}
p^{(1)}_j = \frac{4}{27R_t^4}\int_{\Delta^j_1} Area(\mathcal{H}((x,y),R_{i}) \cap \Delta^j_2) \, dx \, dy 
\end{equation}

Next, we consider various possible interference scenarios for $j^{\textrm{th}}$ tier co-channel cells and compute $p_j$ for each of the cases. \\

\noindent
\textbf{\textit{Case 1: }} $D_j - 2R_t \leq R_{i}  < D_j - R_t$, or equivalently $\nu_j - 2 \leq \gamma(\eta, P_t, P_{min})  < \nu_j-1$. In this scenario, we can rewrite integral \eqref{eq:integral_final} as
\begin{equation} \label{eq:integral_case_1}
p^{(1)}_j = \frac{4}{27R_t^4}\int_{\Delta} Area(\mathcal{H}((x,y),R_{i}) \cap \mathcal{H}((x_2,y_2),R_{t})) \, dx \, dy 
\end{equation}

\begin{figure}[ht]
\centering
\includegraphics[scale=0.55]{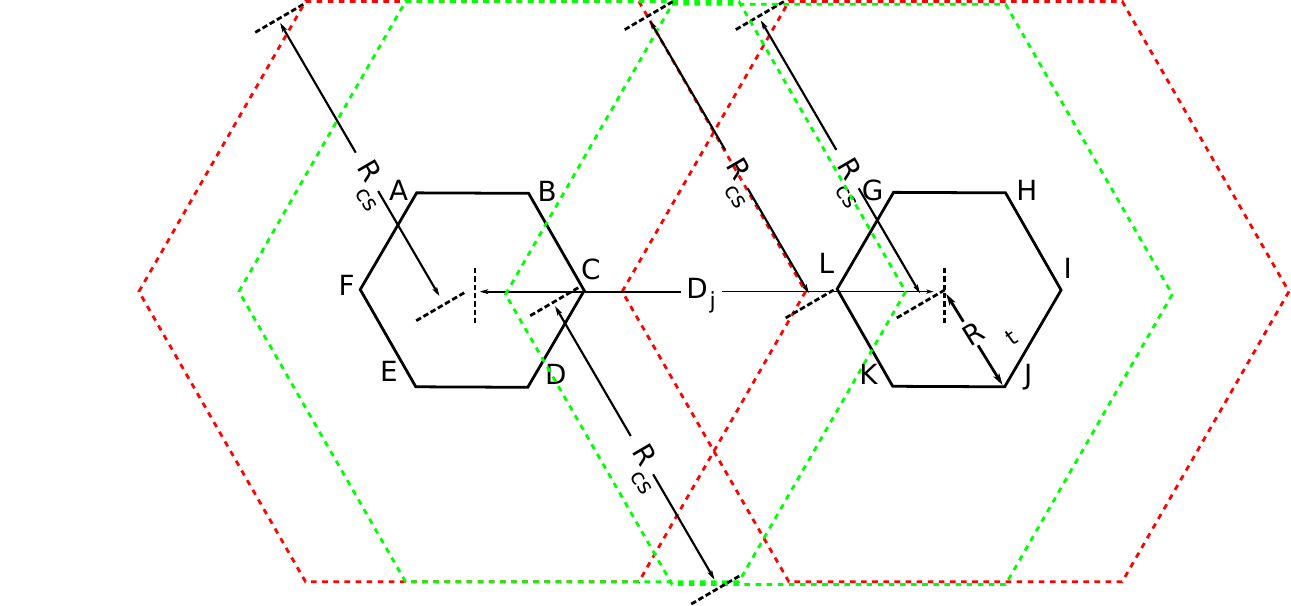}
\caption{Possible scenario for type \Rmnum{1} dependency under Case 1. Red and green hexagons show the interference range of APs and STAs respectively.}
\label{fig:case_1}
\end{figure}

\noindent
where $\Delta=\mathcal{H}((x_1,y_1),R_{t}) \cap \mathcal{H}((x_2-R_t,y_2),R_{t})$. After some geometric constructions, integral \eqref{eq:integral_case_1} becomes
\begin{eqnarray*}
p^{(1)}_j &=& \frac{2}{9 \sqrt{3} R_t^4}\int^{R_{i} + 2R_t - D_j}_{0} \int^{R_{i} + 2R_t - D_j}_{0} xy  \, dx \, dy  = \frac{1}{18 \sqrt{3}} \cdot \left(\frac{R_{i} - D_j + 2R_t}{R_t} \right)^4 
\end{eqnarray*}
Substituting for $R_{i}$ and $D_j$ in terms of $R_t$, we obtain
$$p^{(1)}_j = \frac{1}{18\sqrt{3}} \cdot \left(  \gamma(\eta, P_t, P_{min}) + 2 - \nu_j \right)^4 $$

\noindent
\textbf{\textit{Case 2: }} $D_j - R_t\leq R_{i}  < D_j$, or equivalently $\nu_j - 1 \leq \gamma(\eta, P_t, P_{min})  < \nu_j$ 

\begin{figure}[ht]
\centering
\includegraphics[scale=0.55]{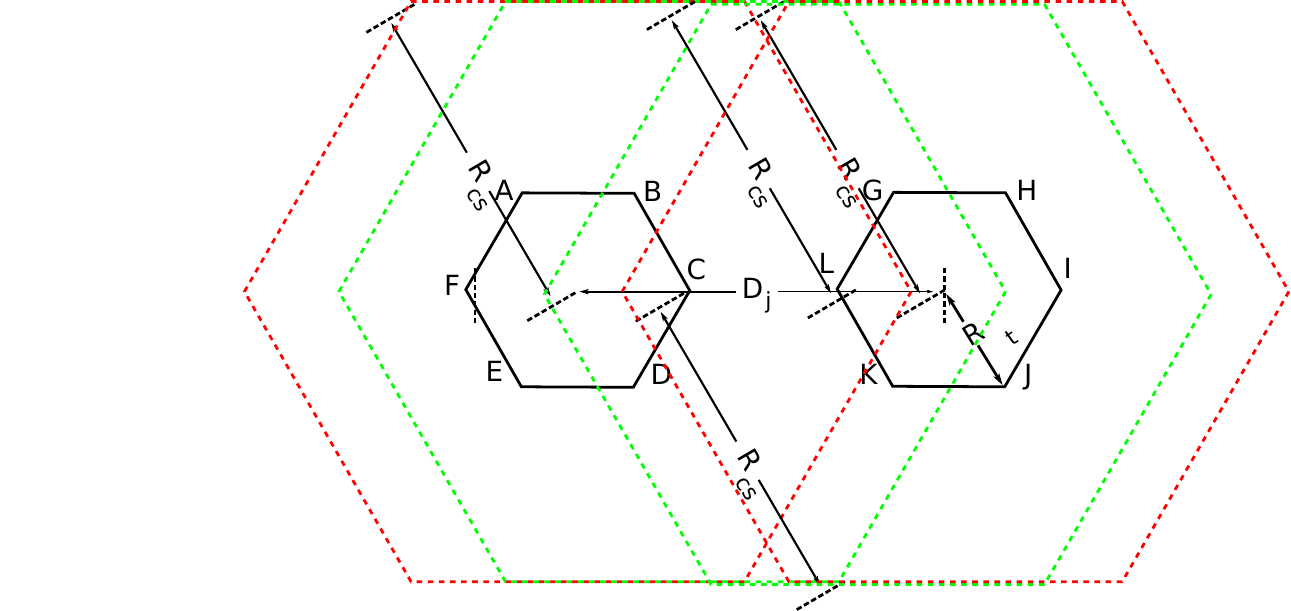}
\caption{Possible scenario for type \Rmnum{1} dependency under Case 2. Red and green hexagons show the interference range of APs and STAs respectively.}
\label{fig:case_2}
\end{figure}

For this case, we evaluate integral \eqref{eq:integral_final} by splitting $\Delta^j_1$ and $\mathcal{H}((x,y),R_{i}) \cap \Delta^j_2$ into non-overlapping area. After some inferences based on geometry and calculus, we get
\begin{eqnarray*}
p^{(1)}_j = \frac{1}{9 \sqrt{3}} \cdot \left( 1 - \left(\frac{R_{i} - D_j + R_t}{R_t}\right)^4  - \frac{1}{2} \cdot \left(\frac{D_j - R_{i}}{R_t}\right)^4 \right) + \frac{1}{54} \cdot \left(\frac{R_{i} - D_j + R_t}{R_t}\right)^4
\end{eqnarray*}
Substituting for $R_{i}$ and $D_j$ in terms of $R_t$, we obtain
\begin{eqnarray*}
p^{(1)}_j = \frac{1}{9 \sqrt{3}} \cdot \left( 1 - \left(\gamma(\eta, P_t, P_{min}) + 1 - \nu_j\right)^4  - \frac{1}{2} \cdot \left( \nu_j - \gamma(\eta, P_t, P_{min}) \right)^4 \right) + \frac{1}{54} \cdot \left( \gamma(\eta, P_t, P_{min}) + 1 - \nu_j\right)^4
\end{eqnarray*}

\noindent
\textbf{\textit{Case 3: }} $D_j \leq R_{i}$, or equivalently $\nu_j \leq \gamma(\eta, P_t, P_{min})$. In this case, the access points are within interference range of each other. Thus, in this scenario, it is impossible to have just \emph{STA-STA} dependency between stations $S_1$ and $S_2$. Thus, we have $p^{(1)}_j = 0$

\subsection{Probability of Type \Rmnum{2} dependency} \label{sec:type_2}
In this section, we find the probability of type \Rmnum{2} dependency between two station $S_1$ and $S_2$. Let $p^{(2)}_j$ denote the probability that two stations have type \Rmnum{2} dependency, given that they belong to co-channels cells with centres  $D_j$ unit apart. The computation of type \Rmnum{2} dependence probability can be split into two simple cases as below. 

\noindent 
\textbf{\textit{Case 1: }} $D_j - 2R_t \leq R_{i}  < D_j$, or equivalently $\nu_j - 2 \leq \gamma(\eta, P_t, P_{min})  < \nu_j$.
In this case, the co-channel access points are out of each others interference range. Thus, $p^{(2)}_j = 0$

\noindent
\textbf{\textit{Case 2: }} $D_j \leq R_{i}$, or equivalently $\nu_j \leq \gamma(\eta, P_t, P_{min})$.
In this case, the co-channel access points are within each others interference range. Thus, $p^{(2)}_j = 1$

\subsection{Probability of Type \Rmnum{3} dependency} \label{sec:type_3}
In this section, we find the probability of type \Rmnum{3} dependency between two station $S_1$ and $S_2$. Let $p^{(3)}_j$ denote the probability that two stations have type \Rmnum{3} dependency, given that they belong to co-channels cells with centres  $D_j$ unit apart. 

\noindent
\textbf{\textit{Case 1: }} $D_j - 2R_t \leq R_{i}  < D_j - R_t$, or equivalently $\nu_j - 2 \leq \gamma(\eta, P_t, P_{min})  < \nu_j - 1$. In this case, the interference region of the access points do no over lap the $j^{\textrm{th}}$ tier co-channel cells. Thus, in this case, type \Rmnum{3} dependency cannot occur i.e., $p^{(3)}_j = 0$

\noindent
\textbf{\textit{Case 2: }} $D_j - R_t \leq R_{i}  < D_j$, or equivalently $\nu_j - 1 \leq \gamma(\eta, P_t, P_{min})  < \nu_j$. For this case, we have
\begin{eqnarray*}
p^{(3)}_j = 1 - P[\textrm{Access point } A_1 \textrm{ does not interfere with station } S_2 \textrm{ and access point } A_2 \textrm{ does interferes with station } S_1]	
\end{eqnarray*}
From application of simple geometric argument, we get 
\begin{eqnarray*}
p^{(3)}_j = 1 - \left(1 - \frac{\sqrt{3} \cdot (R_i + R_t - D_j)^2/2}{3\sqrt{3} R^2_t/2} \right)^2 = 1 - \left(1 - \frac{1}{3} \cdot (\gamma(\eta, P_t, P_{min}) + 1 - \nu_j)^2 \right)^2
\end{eqnarray*}

\noindent
\textbf{\textit{Case 3: }} $D_j \leq R_{i}$, or equivalently $\nu_j  \leq \gamma(\eta, P_t, P_{min})$. In this case, the access points interfere with each other. Thus, we do not have type \Rmnum{3} dependency i.e., $p^{(3)}_j = 0$

\subsection{Final probability expression for each type of dependency}
Let $\mathcal{N}_j \subset \mathcal{N}$ denote the $j^{\textrm{th}}$ tier co-channel cells of cell $d$ and let $n_j$ be its cardinality. Let $\mathcal{N}^{'} = \cup^{j_0}_{j = 1} \mathcal{N}_j$. Since elements of the set $\{ \mathcal{N}_j, j_0 \geq j \geq 1 \}$ do not overlap with each other, we have $|\mathcal{N}^{'}| = \sum_{j_0 \geq j \geq 1} n_j$. Let us define an indicator variable as follows:
$$I^j(S_2)  = 
\begin{cases}
1 & \textrm{ if station } S_2 \textrm{ is in a } j^{\textrm{th}} \textrm{ tier co-channel cell}\\
0 & \textrm{ otherwise }
\end{cases}$$
Let $E^{(i)}$ and $p^{(i)}$ denote the event and probability of type $i$ dependency between stations $S_1$ and $S_2$ given that $S_2 \in \mathcal{N}^{'}$. Then, we have
\begin{eqnarray*}
p^{(i)}  = P[E^{(i)} | S_2 \in \mathcal{N}^{'}] = \sum_{l \in \mathcal{N}^{'}} P[S_2 \in Cell \, l  , E^{(i)} | S_2 \in \mathcal{N}^{'}] 
= \sum_{l \in \mathcal{N}^{'}} \sum_{j_0 \geq j \geq 1} P[S_2 \in Cell \, l, E^{(i)}, I^{j}(S_2)=1 | S_2 \in \mathcal{N}^{'}]
\end{eqnarray*}
By conditioning on the events and simplifying, we get
\begin{eqnarray} \label{eq:p_final}
p^{(1)}  =  \sum_{l \in \mathcal{N}^{'}} \sum_{j_0 \geq j \geq 1} P[ S_2 \in Cell \, l | S_2 \in \mathcal{N}^{'}] \cdot P[ I^{j}(S_2) = 1| S_2 \in Cell \, l] \cdot P[ E^{(i)} | I^{j}(S_2)=1, S_2 \in Cell \, l]
\end{eqnarray}
By definition, we have 
$$P[ E^{(i)} | I^{j}(S_2)=1, S_2 \in Cell \, l] = p^{(i)}_j \quad  \forall l \in \mathcal{N}^{'}$$ 
Station $S_2$ has a uniform distribution over the deployed area, we have $P[ S_2 \in Cell \, l | S_2 \in \mathcal{N}^{'}] =  \frac{1}{|\mathcal{N}^{'}|}, \forall l \in \mathcal{N}^{'}$. We also have 
$$P[ I^{j}(S_2) = 1| S_2 \in Cell \, l] = 
\begin{cases}
1 & \textrm{ if } S_2 \in \mathcal{N}_j \\
0 & \textrm{ otherwise }
\end{cases} $$
Plugging in the various probabilities into Equation \eqref{eq:p_final}, we obtain 
\begin{eqnarray*} \label{eq:p_final_1}
p^{(i)}  =  \sum_{j_0 \geq j \geq 1} p^{(i)}_j \cdot P[S_2 \in \mathcal{N}_j | S_2 \in \mathcal{N}^{'}] = \left(\frac{1}{\sum_{j_0 \geq j \geq 1} n_j}\right) \cdot \sum_{j_0 \geq j \geq 1} p^{(i)}_j  \cdot n_j
\end{eqnarray*}
where $n_j \in \{6, 12 \}$ is the number of $j^{\textrm{th}}$ tier co-channel cells of cell $d$.

\begin{table}[ht]
\centering
\caption{Table showing the values of $p$ against various values of $r_t$, given that $\alpha = 1 \textrm{ and } \eta = 3.5$}	
\begin{tabular}{|c|c|c|c|c|}
\hline 
$r_t$ & $P_t$ & $p^{(1)}$ & $p^{(2)}$ & $p^{(3)}$ \\ 
\hline 
 54 & -44 & 0.0010 & 0.9510 & 0.0112 \\ 
\hline 
 48 & -60 & 0.0074 & 0.7778 & 0.0112 \\ 
\hline 
 36 & -69 & 0.0076 & 0.7143 & 0.0000 \\ 
\hline 
 24 & -73 & 0.0000 & 0.6000 & 0.0000 \\ 
\hline 
 12	 & -85 & 0.0416 & 0.0000 & 0.3686 \\ 
\hline 
 1 & -90 & 0.0321 & 0.0000 & 0.0000  \\ 
 \hline
\end{tabular} 
\end{table}

\end{document}